\documentclass{aastex} 
\usepackage{emulateapj5}
\usepackage{times,mathptmx}

\slugcomment{Submitted to ApJ}

\shorttitle{Stellar masses of field galaxies in MUNICS}
\shortauthors{Drory et al.}

\newcommand*{\Msun}{\ensuremath{\mathrm{M_\odot}}}%
\newcommand*{\hMsun}{\ensuremath{h^{-2}\,\mathrm{M_\odot}}}%
\newcommand*{\Mlim}{\ensuremath{M_{\mathrm{lim}}}}%
\newcommand*{\MLK}{\ensuremath{M/L_K}}%
\newcommand*{\MLB}{\ensuremath{M/L_B}}%
\newcommand*{\ML}{\ensuremath{M/L}}%
\newcommand*{\Vmax}{\ensuremath{V_{\mathrm{max}}}}%
\newcommand*{\rhos}{\ensuremath{\rho_*}}%
\newcommand*{\rhosz}{\ensuremath{\rho_*(z)}}%
\newcommand*{\Mpc}{\ensuremath{\mathrm{Mpc}}}%
\newcommand*{\E}[1]{\ensuremath{\times 10^{#1}}}%
\newcommand*{\p}{\ensuremath{\pm}}%

\begin{document}


\title{The Munich Near--Infrared Cluster Survey (MUNICS) -- VI. The
  stellar masses of K--band selected field galaxies to $z~\sim~1.2$}


\author{N.~Drory\altaffilmark{1}, R.~Bender\altaffilmark{2,3},
  G.~Feulner\altaffilmark{2}, U.~Hopp\altaffilmark{2},
  C.~Maraston\altaffilmark{3}, J.~Snigula\altaffilmark{2}} \and
\author{G.~J.~Hill\altaffilmark{1}}

\affil{$^1$ University of Texas at Austin, Austin, Texas 78712}
\email{\{drory,hill\}@astro.as.utexas.edu}

\affil{$^2$ Universit\"ats--Sternwarte M\"unchen, Scheinerstra\ss
  e 1, D-81679 M\"unchen, Germany}
\email{\{bender,feulner,hopp,snigula\}@usm.uni-muenchen.de}

\affil{$^3$ Max--Planck Institut f\"ur extraterrestrische Physik,
  Giessenbachstra\ss e, Garching, Germany}
\email{\{bender,maraston\}@mpe.mpg.de}


\begin{abstract}
  We present a measurement of the evolution of the stellar mass
  function in four redshift bins at $0.4 < z < 1.2$, using a sample of
  more than 5000 $K$-selected galaxies drawn from the MUNICS (Munich
  Near-Infrared Cluster Survey) dataset.  Our data cover the stellar
  mass range $10^{10} \leq M/(\hMsun) \leq 10^{12}$. We derive K--band
  mass--to--light ratios by fitting a grid of composite stellar
  population models of varying star formation history, age, and dust
  extinction to BVRIJK photometry. We discuss the evolution of the
  average mass--to--light ratio as a function of galaxy stellar mass
  in the K and B bands.  We compare our stellar mass function at $z >
  0$ to estimates obtained similarly at $z=0$. We find that the
  mass--to--light ratios in the K--band decline with redshift. This
  decline is similar for all stellar masses above $10^{10}\,\hMsun$.
  Lower mass galaxies have lower mass--to--light ratios at all
  redshifts. The stellar mass function evolves significantly to $z =
  1.2$. The total normalization decreases by a factor of $\sim 2$, the
  characteristic mass (the knee) shifts toward lower masses, and the
  bright end therefore steepens with redshift. The amount of number
  density evolution is a strong function of stellar mass, with more
  massive systems showing faster evolution than less massive systems.
  We discuss the total stellar mass density of the universe and
  compare our results to the values from the literature at both lower
  and higher redshifts.  We find that the stellar mass density at $z
  \sim 1$ is roughly 50\% of the local value. Our results imply that
  the mass assembly of galaxies continues well after $z \sim 1$. Our
  data favor a scenario in which the growth of the most massive
  galaxies is dominated by accretion and merging rather than star
  formation which plays a larger role in the growth of less massive
  systems.
\end{abstract}

\keywords{surveys --- cosmology: observations --- galaxies: mass
function --- galaxies: evolution --- galaxies: fundamental parameters}


\section{Introduction}\label{sec:introduction}

The stellar mass content in galaxies as a function of redshift is one
of the most fundamental observables in the quest to understand galaxy
formation and evolution. It provides information on the coupling
between the growth of structure through the collapse and subsequent
merging of dark matter halos and the physical processes governing the
evolution of the baryonic matter. Stellar mass in galaxies grows by
star formation within galactic disks, as well as by accretion and the
merging of galaxies.  In fact, these two processes are related,
because star formation in disks can be triggered or enhanced by tidal
interaction in close encounters and by merging events.  This interplay
between cosmological structure formation and star formation is
believed to govern the mass assembly history of galaxies.

The stellar mass of a galaxy at a given time is difficult to measure,
however. While dynamical mass measures are considered to be most
reliable, they measure the total mass of an object. The dark matter
and gas contributions (which are a function of galaxy type) have to be
removed to obtain the stellar mass. These kinds of measurements depend
on model assumptions for the dark matter contribution, are
observationally very costly, and have therefore only been possible in
the local universe so far.

The alternative is to convert the luminosity of a galaxy into a
stellar mass by means of a model of its stellar population (derived
from photometry or spectroscopy) predicting a mass--to--light ratio
(\ML) in a certain wavelength band.  Near--infrared (NIR) luminosities
of galaxies are believed to be well suited for this approach, as the
\ML values vary only by a factor of roughly 2 across a wide range of
star formation histories (SFHs; see, e.g., \citealp{RR93,KC98a,BD01}).
This compares to a variation of a factor of $\sim 10$ in the B--band.
In addition, the optical regime is strongly affected by dust
extinction which becomes negligible in the K band for the vast
majority of galaxies \citep{TPHSVW98}. By correlating photometric
properties of disk galaxies with inclination, \citet{MGH03} found the
edge--on to face--on extinction correction to be 0.1~mag in the K
band.

Measuring the stellar masses of galaxies in the local universe by
means of modeling their stellar populations has been re--attempted
recently using newly available wide--area galaxy surveys.
\citet{Kauffmannetal03a} used spectroscopic data from the Sloan
Digital Sky Survey (SDSS), while \citet{2dF01} and \citet{BMKW03}
combined NIR photometry from the Two Micron All Sky Survey (2MASS)
with optical photometry from the 2dF Galaxy Redshift Survey (2dFGRS)
and SDSS, respectively, to derive \ML values and study the stellar
mass function (MF) of galaxies.

At $z > 0$, suitable multi--wavelength and redshift data are still
sparse. Therefore, the integrated stellar mass density, \rhosz, has
been studied, instead of the stellar MF using the available deep field
observations. \citet{DPFB03} and \citet{Fontanaetal03} studied \rhosz\ 
in the Hubble Deep Fields (HDFs) over the redshift range $0<z<3$ and
compared the global integrated SFH of the universe to their
measurement of the total stellar mass density.  These surveys,
however, cover only very small fields (several square arcminutes) and
therefore cannot measure the evolution of \rhosz\ below $z\sim 1$.  In
addition, at higher $z$, cosmic variance, selection biases, and dust
extinction are a concern, as the ultraviolet spectral region has
redshifted into the observable bands.  At $0<z<1$, the total stellar
mass density has been estimated from optical and NIR surveys
(\citealp{BE00}; \citealp{MUNICS3}, hereafter MUNICS~III;
\citealp{Cohen02}), yet the results concerning the evolution of
\rhosz\ in this redshift range are still inconclusive.  New
wide--angle, NIR selected surveys are promising rapid progress
(\citealp{K20-03,LCIRS02};\citealp{MUNICS1}, hereafter MUNICS~I) in
studying the stellar MF itself instead.

In this work we use data from the wide--field K-band selected field
galaxy sample of the MUNICS project (MUNICS~I; \citealp{MUNICS5},
hereafter MUNICS~V) to study the stellar MF evolution to $z \sim 1.2$
directly. We derive stellar masses by converting K-band luminosities
to mass, by comparing BVRIJK photometry of galaxies to a grid of
composite stellar population (CSP) synthesis models based on the
simple stellar populations published by \citet{Maraston98}.

We introduce the sample used in this work in
Sect.~\ref{sec:galaxy-sample} and the method used to derive stellar
masses and its uncertainties in Sect.~\ref{sec:deriving-masses}. We
present the results for the stellar MF in
Sect.~\ref{sec:mass-function} and discuss the number density of
galaxies with $10^{10} \leq M/(\hMsun) \leq 10^{12}$ in
Sect.~\ref{sec:num-density}. We present our estimate of the total
stellar mass density at $0.4 < z < 1.2$ and compare it to those in the
literature in Sect.~\ref{sec:mass-density}. Finally, we summarize and
discuss our results in Sect.\ref{sec:discussion}.

Throughout this work we assume $\Omega_M = 0.3$, $\Omega_{\Lambda} =
0.7$. We write Hubble's constant as $H_0 = 100\ h\ \mathrm{km\ s^{-1}\ 
  Mpc^{-1}}$, unless noted otherwise. We denote absolute magnitudes in
band $X$ by the symbol $M_X$ and masses by the symbol $M$.


\section{The galaxy sample}\label{sec:galaxy-sample}

MUNICS is a wide--area, medium--deep, photometric and spectroscopic
survey selected in the K band, reaching $K \sim 19.5$. It covers an
area of roughly one deg$^2$\ in the K and J bands with optical
follow--up imaging in the I, R, V, and B bands in 0.4 deg$^2$.
MUNICS~I discusses the field selection, object extraction, and
photometry. Detection biases, completeness, and photometric biases of
the MUNICS data are analyzed in detail in \citet[][hereafter
MUNICS~IV]{MUNICS4}.

The MUNICS photometric survey is complemented by spectroscopic
follow--up observations of all galaxies down to $K\le 17.5$ in 0.25
deg$^2$, and a sparsely selected deeper sample down to $K \le 19$.  It
contains 593 secured redshifts to date. The spectra cover a wide
wavelength range of $4000-8500$\AA\ at $13.2$\AA\ resolution, and
sample galaxies at $0<z<1$. These observations are described in detail
in MUNICS~V.

The galaxy sample used in this work is a subsample of the MUNICS
survey Mosaic Fields (see MUNICS~I), selected for best photometric
homogeneity, good seeing, and similar depth. The subsample covers 0.28
deg$^2$\ in the B, V, R, I, J, and K bands. It is identical to the
sample used in \citet[][hereafter MUNICS~II]{MUNICS2} to derive the
evolution of the K--band luminosity function (LF). MUNICS~II also
discusses the procedure used to obtain photometric redshifts for the
full sample and the calibration of the technique using spectroscopic
redshifts. The reader is referred to this paper for details.  The
sample used in MUNICS~III to derive number densities as a function of
mass is selected from the same survey areas but lacks B-band
photometry which was added to MUNICS at a later time.


\section{Deriving stellar masses}\label{sec:deriving-masses}

Measuring stellar masses by estimating \ML\ values in distant galaxies
poses some methodological difficulties. At the bright end, even small
errors in \ML\ translate into large errors in the resulting stellar
mass, since the LF is very steep. At the faint end, the dynamic range
in \ML\ values increases, and mean stellar ages are harder to derive
because of the possibly more complicated SFHs and larger mass
fractions of stars born in recent bursts.  Therefore, it is important
to carefully evaluate the procedure that is used.

In our case, we have photometry in six pass--bands (five colors).
Deriving an \ML\ for each object requires us to estimate a number of
parameters similar to the number of observables.  We need a
photometric redshift (although in MUNICS this is achieved
independently of the CSP models used here, it involves the same
photometry), an SFH, a mean stellar age, and the amount of dust
extinction. These are already four free parameters, and we therefore
restrict ourselves to this minimal set of models, leaving out
metallicity and superimposed bursts of star formation, instead of
including more parameters and marginalizing over them. As noted below,
the addition of bursts of star formation has much smaller effects on
the NIR \ML\ values than it has on the optical ones.  Since the MUNICS
sample contains galaxies more massive than $\sim 10^{10} \Msun$, it is
unlikely that many systems with low metallicity are present in the
sample, and we may restrict ourselves to solar metallicity models,
especially since the \ML\ estimates are rather robust with respect to
a changes in metallicity around the solar value (see also the
discussion of uncertainties in \ML\ below).

A possible concern might be that we fit spectral energy distributions
(SEDs) twice, using one set to derive a photometric redshift and
another set to estimate stellar masses, and taht those two sets might
not be independent. However, it is important to point out that since
we use a much smaller set of semi-empirical SEDs (combining observed
SEDs and models) to obtain photometric redshifts, the SEDs used here
and those used for the photometric redshifts are, in fact, largely
independent. The photometric redshift code uses a set of SEDs that are
free combinations of observed galaxy SEDs and model SEDs fitted to
combined broad band photometry of objects with spectroscopic redshifts
as described in MUNICS~II. Hence, they do not allow straight-forward
interpretation in terms of physical parameters, such as age or SFH
(and, in fact, need not even cover this parameter space in any
meaningful way, only the observed colors of galaxies as a function of
redshift), and we need an independent grid covering the physical
parameter space for the present analysis.

We parameterize the SFH as $\psi(t) \propto \exp(-t/\tau)$, \\ with
$\tau \in \{0.1, 0.2, 0.4, 1.0, 2.0, 3.0, 5.0, 8.0, 10.0, 13.0\}$~Gyr.
We extract spectra at 28 ages between 0.001 and 14~Gyr and allow $A_v$
to vary between 0 and 3 mag, using a \citet{Calzetti00} extinction
law.  The models use solar metallicity and are based on the simple
stellar population models by \citet{Maraston98}. We assume a value of
3.33 for the absolute K--band magnitude of the Sun. We use a Salpeter
initial mass function (IMF), with lower and upper mass cutoffs of 0.1
and 100 \Msun. This choice of IMF allows us to compare our results to
those in the literature more easily. The use of an IMF with a flatter
slope at the low mass end will not affect the shape of the MF, it will
only change its overall normalization. If, however, the IMF depends on
the mode of star formation, e.g.\ being top--heavy in starbursts, our
results will be affected. This particular choice of model grid
parameters yields a fairly uniform coverage of color space and
represents the SEDs of galaxies in the sample reasonably well in a
$\chi^2$ sense.

We convert the absolute K--band magnitude, $M_K$, into stellar mass by
using the K--band mass--to--light ratio, \MLK, of the best fitting CSP
model in a $\chi^2$\ sense. We weight by the errors of the photometry
and assume an uncertainty of 5\% in all model colors to account for
the discreteness of the model grid. In addition, another 5\%
uncertainty is added at J and K wavelengths to account for the
intrinsic uncertainty of the models at NIR wavelengths. We employ an
age prior falling off as $\exp\{(-\Delta t/1~\mathrm{Gyr})^4\}$ for
ages $\Delta t = t - t_H(z) > 0$, $t_H$ being the age of the universe
at redshift $z$, to suppress models with ages greater than the age of
the universe at any given redshift.

The value of $M_K$ is obtained by taking the restframe K--band
magnitude of the best fitting SED of the photometric redshift code of
each object (the identical magnitudes were used to construct the LF;
see MUNICS~II). Since the SED is chosen by a fit using all six
pass--bands, the extrapolation to restframe $K$ is based on the same
information as an interpolation to restframe $J$ or any other
wavelength using this SED would be, and it is therefore no worse.
Because NIR SEDs do have some broad features and curvature in the
continuum, a simple interpolation between observed magnitudes will not
suffice.

\begin{figure}
  \centering 
  \epsscale{0.6}
  \plotone{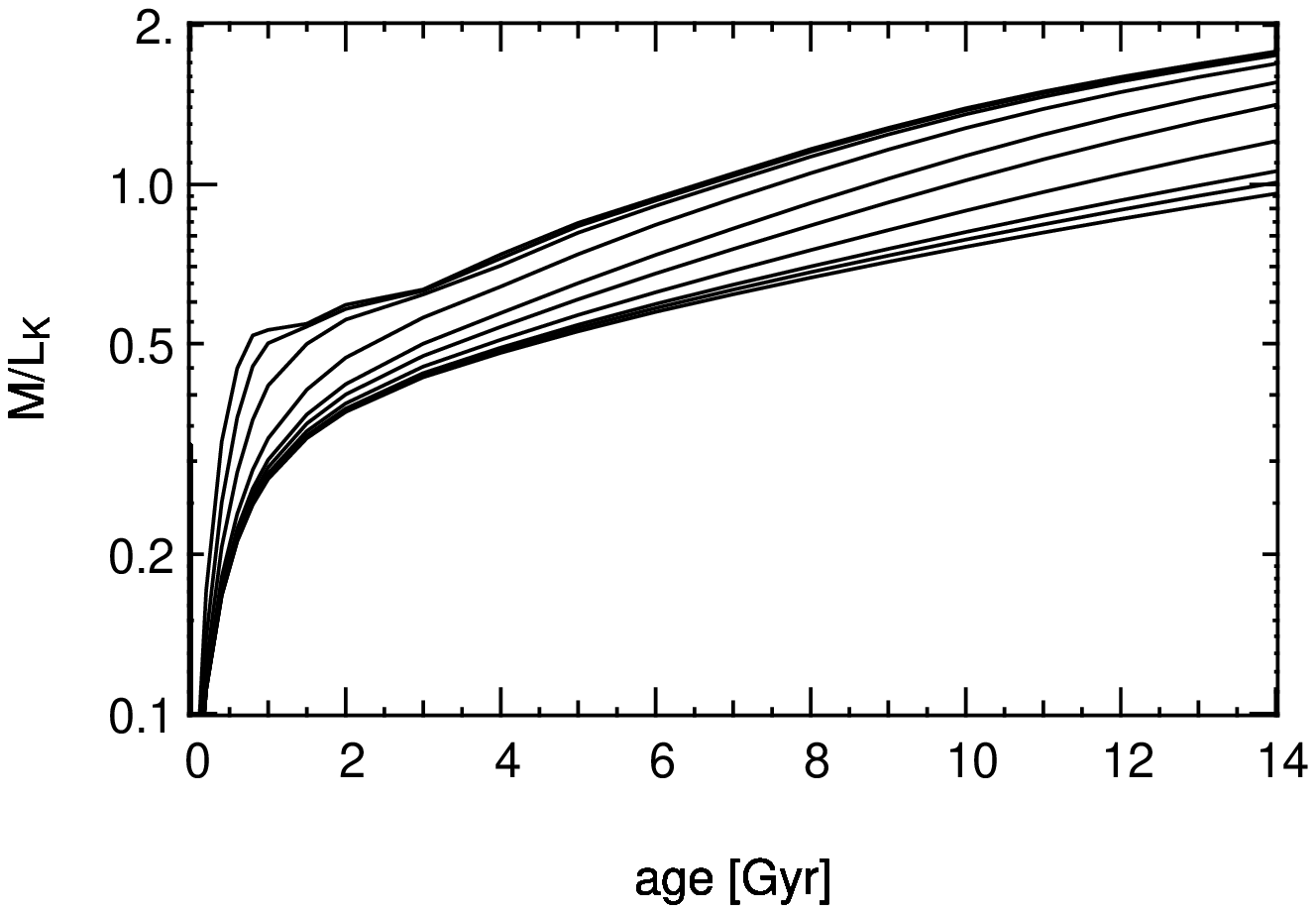}
  \caption{\label{fig:mlk-csps}%
    K--band \ML\ value as a function of age of the CSP
    models that we use.  Lines from top to bottom show \MLK\ of star
    formation histories of the form $\psi(t) \propto \exp(-t/\tau)$,
    with $\tau \in \{0.1, 0.2, 0.4, 1.0, 2.0, 3.0, 5.0, 8.0, 10.0,
    13.0\}$.}
\end{figure}

The \MLK\ values of our CSP models as a function of age are shown in
Fig.~\ref{fig:mlk-csps}.  At ages above $\sim 1.5$~Gyr, the variation
with age of \MLK\ is of the order of a factor of $\sim 3$, while the
variation between the models at any given age is less than a factor of
$\sim 2$. This narrow dynamic range in \ML, along with the negligible
influence of dust extinction, constitutes the advantage of using the
K--band to derive stellar \ML\ values, making the derived masses quite
robust.

We estimate the total systematic uncertainty in the estimated \MLK\ 
for each object due to the limited range in parameter space covered by
the model grid to be roughly 25\% - 30\%. In particular, three sources
of systematic uncertainty contribute to this number: the effect of
neglecting metallicity is found to contribute $\sim 10\%$. We arrive
at this number by comparing model \ML\ values at different ages and
metallicities but similar colors, which are compatible with the
observations within the photometric errors (the well known
age--metallicity degeneracy). This effect is found to be roughly
symmetric with respect to metallicity and therefore, if the average
metallicity of the galaxies in the current sample is close to the
solar value, contributes a close to random uncertainty.  The effect of
starbursts on top of our smooth SFHs is estimated to change \MLK\ to
lower values by 5\% - 10\%. This number is obtained by adding bursts
of up to 10\% in mass and ages of up to 3~Gyr to the $\tau = 5$~Gyr
and $\tau = 10$~Gyr models (assuming no dust extinction; see also
\citealp{BD01} for further discussion of stellar population models of
disk galaxy \ML\ values). Finally, systematic uncertainties in the
colors and \MLK\ values of the underlying stellar population model are
estimated to contribute another 5\% to 10\% uncertainty, especially at
younger ages where supergiants contribute a large fraction of the
K--band light (see also the discussion of the calibration of these
models in \citealp{Maraston98}). The latter would bias the derived
\ML\ to higher values, since the models will tend to underpredict the
light contribution of young populations in the NIR.

Another concern lies in the Kron--like aperture photometry that is
used in the MUNICS survey to measure total magnitudes. We have
performed extensive simulations in MUNICS~IV to evaluate the
reliability of these total magnitude measurements using an empirical
effective surface brightness--size relationship for disk galaxies and
the fundamental plane relation for elliptical galaxies.

The main result from this effort is that generally, the magnitudes of $L^*$
objects are fairly well recovered. Pure exponential profiles suffer
only very little bias and almost no magnitude dependent trend over a
wide range of luminosities, even at the highest redshift we probe
here.  In contrast, de Vaucouleurs profiles show higher lost--light
fractions at bright intrinsic magnitudes of $L \ge L^* + 1$ and a
strong magnitude dependent increase of the lost--light fraction with
increasing intrinsic magnitude. This is due to the fact that brighter
elliptical galaxies have lower mean surface brightness.

\begin{figure}
  \centering 
  \epsscale{0.6}
  \plotone{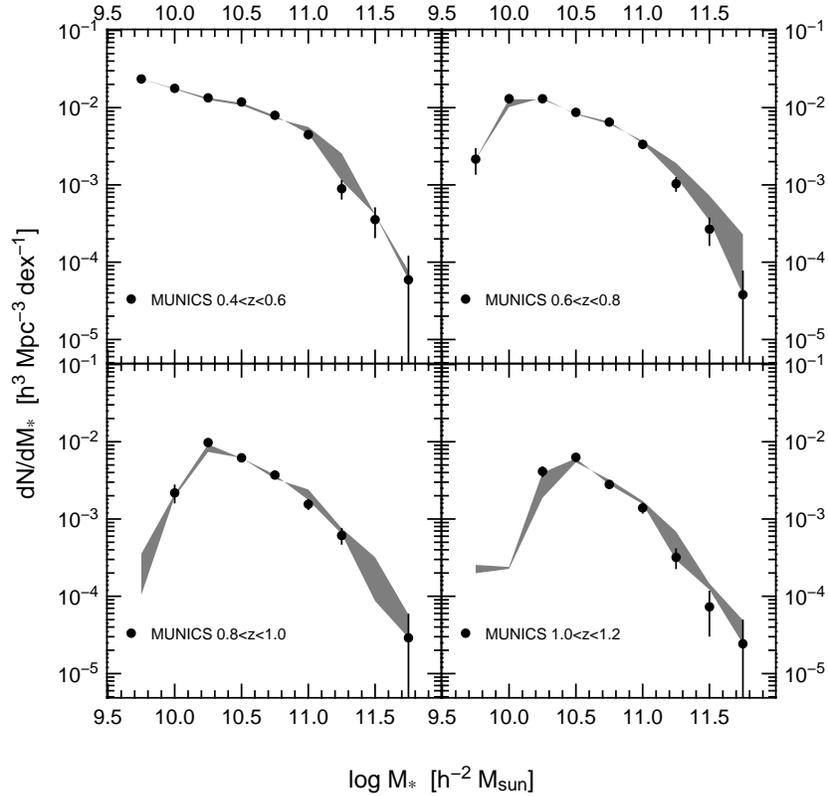}
  \caption{\label{f:mf-llf}%
    The effect of correcting for the lost--light fraction in
    Kron--like magnitudes at high redshift. The solid dots represent
    the raw uncorrected data, the dashed region shows the lost--light
    correction according to simulations in MUNICS~IV. The region is
    bounded by using face on exponential disks and pure de Vaucouleurs
    profiles (see text).}
\end{figure}

We cannot correct for this effect without assuming a morphological mix
of galaxies as a function of redshift or being able to reliably
measure bulge--to--disk ratios in our sample.  Therefore, we calculate
the effect it has on our MFs, by assuming first that all galaxies are
exponential disks and second that all galaxies have de Vaucouleurs
profiles, and show the MFs for both cases in Fig.~\ref{f:mf-llf}. The
data points represent the uncorrected MF and the shaded area the range
of possible values between the two cases.

Note that the assumption that all galaxies follow de Vaucouleurs
profiles does represent an upper limit, since the lost--light fraction
in this case is higher than in any other. The case in which all
galaxies are assumed to follow exponential profiles does not represent
a lower limit, however. Because of its dependence on intrinsic
magnitude, the effect on the MF is most notable at the massive end but
does not dominate the uncertainty there (these bins typically have
fewer than five objects, and the brightest bin has one object; see the
discussion of uncertainties below).  Furthermore, around the knee of
the MF (at the characteristic mass $M^*$), this correction appears
irrelevant, as the intrinsic magnitudes there are well recovered. It
also seems that the shape of the MF is altered only very little.
Finally, the reader might note that in some cases at the (incomplete)
faint end (e.g., at $z=1.1$ and $\log M = 10.25$) the corrected range
extends below the uncorrected data. This is an artifact caused by the
incomplete data at lower masses and by binning.  Objects in a bin are
corrected toward higher masses (and move out of the bin), while too
few objects are corrected into the bin from below (because of strong
incompleteness). The bin therefore effectively loses objects.


\section{The stellar mass function}\label{sec:mass-function}

\begin{figure*}
  \centering 
  \epsscale{0.3}
  \plotone{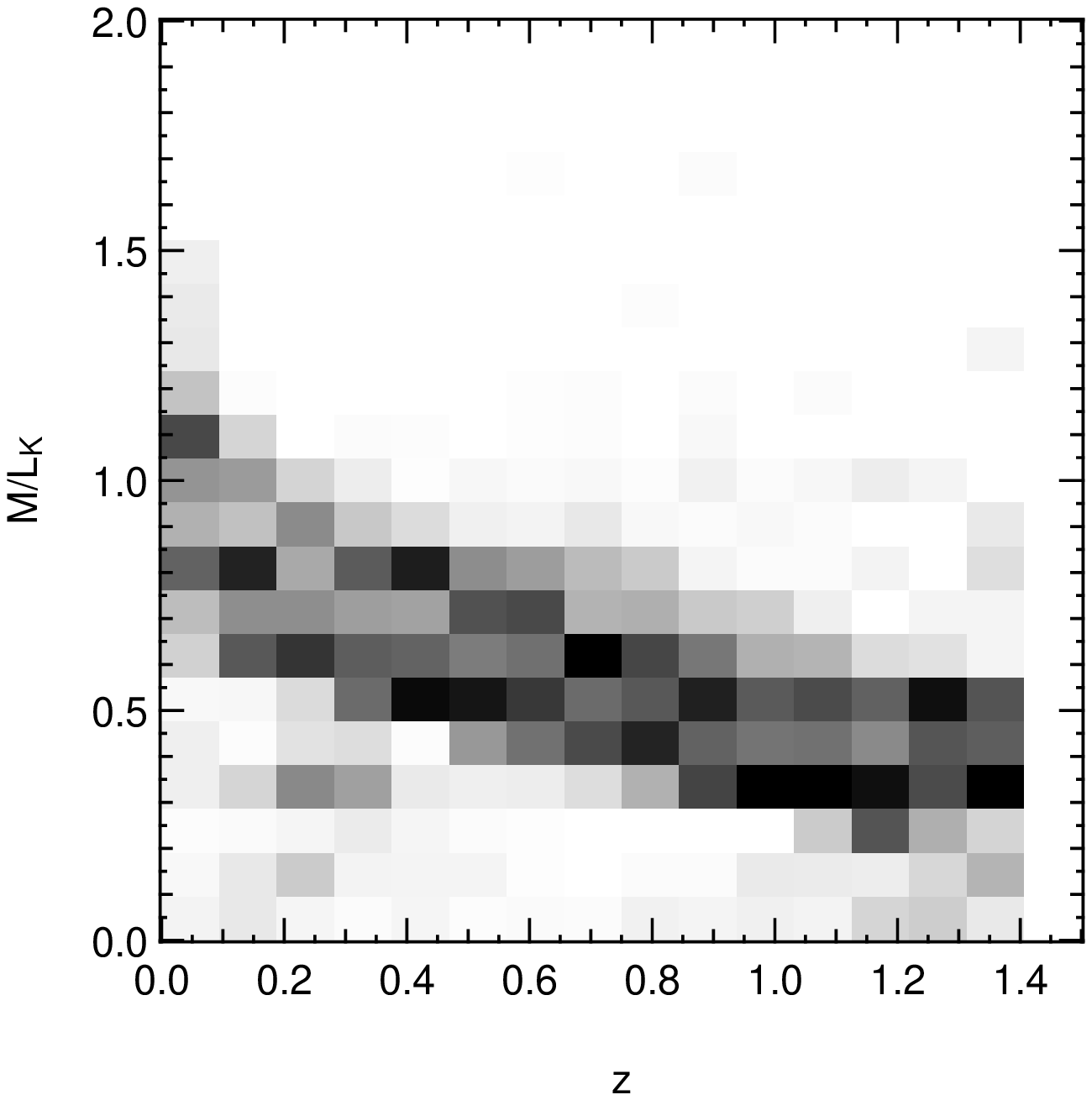}
  \plotone{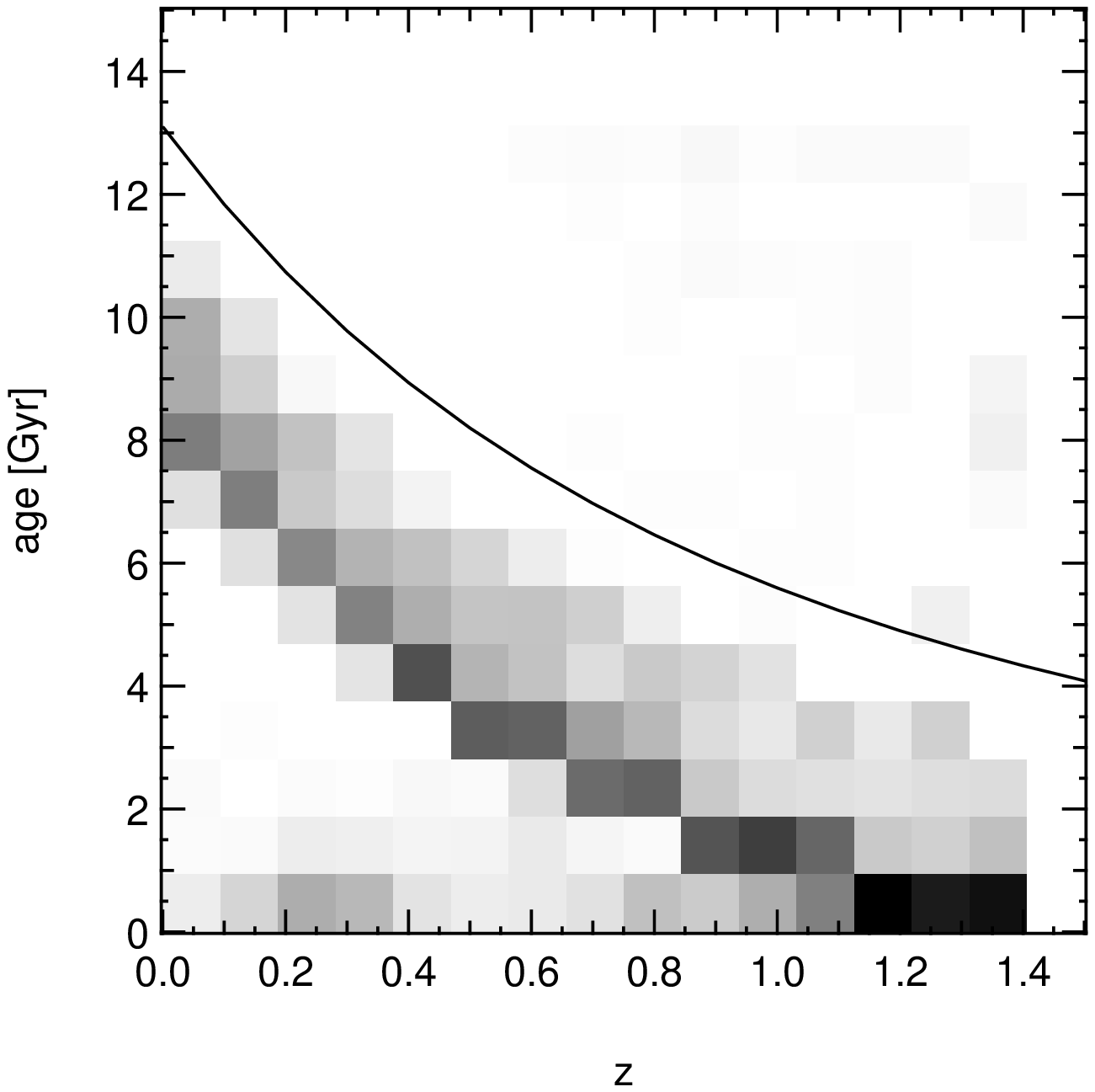}
  \plotone{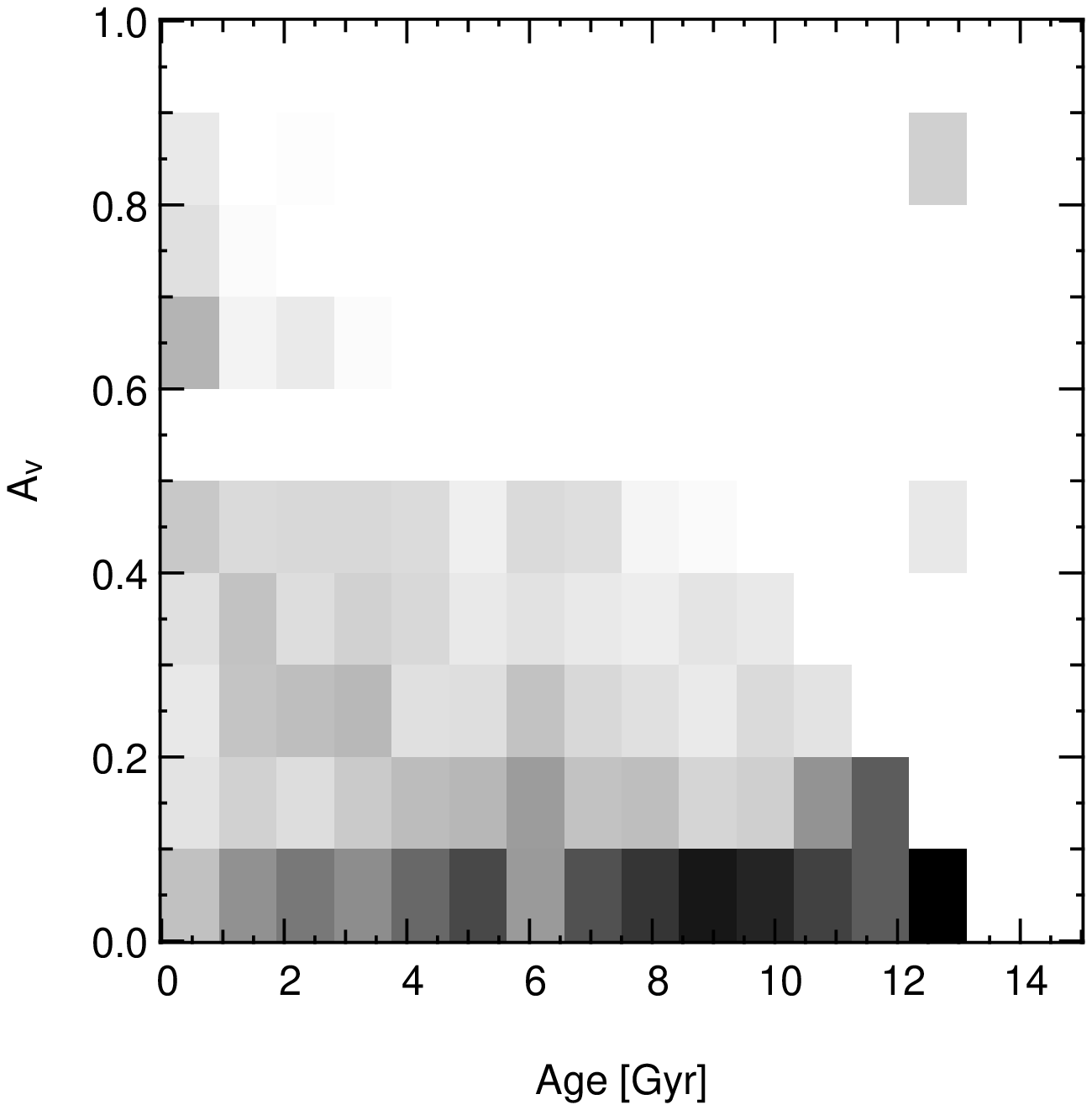}
  \caption{\label{fig:fit-distrib}%
    Distributions of model parameters that we obtain by fitting BVRIJK
    multicolor data to the CSP model grid. The left--hand panel shows
    the distribution of the K--band \ML\ values with
    redshift.  The middle panel the distribution of mean luminosity
    weighted stellar age as a function of redshift, the age of the
    universe using $h=0.72$ is plotted as a reference (solid line).
    The right hand panel shows the distribution of the the dust
    extinction coefficient $A_V$ vs. the luminosity weighted mean
    age.}
\end{figure*}

\begin{figure}
  \centering 
  \epsscale{0.6}
  \plotone{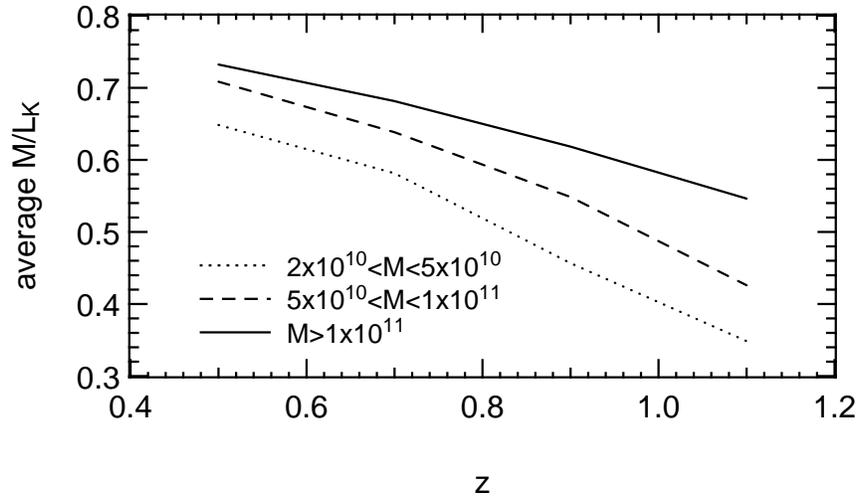}
  \caption{\label{f:mean-mlk-mass}%
    The average K--band \ML\ value as a function of stellar
    mass and redshift.}
\end{figure}

The distributions of model parameters that we obtain by fitting the
data to the CSP model grid described above are shown in
Fig.~\ref{fig:fit-distrib}. The left--hand panel shows the
distribution of the K--band \ML\ values with redshift. The middle
panel shows the distribution of mean luminosity weighted stellar age
as a function of redshift. The age of the universe is plotted as a
reference (solid line), using $h=0.72$.  The right hand panel shows
the distribution of the dust extinction coefficient $A_V$ versus the
luminosity weighted mean stellar age. The average \Vmax--weighted
K--band \ML\ values in our four redshift bins are given in
Table~\ref{tab:mean-mlk-mass} and plotted in
Fig.~\ref{f:mean-mlk-mass}. Table~\ref{tab:mean-mlk-mass} also lists
the average B--band \ML\ values for comparison to other work.

At $z \sim 0$, the mean luminosity weighted stellar age is 8.1~Gyr
with a mean \MLK\ of 0.91. Most objects have \MLK\ between 0.7 and 1.2
and the majority have moderate dust extinction of $A_V < 0.6$. Note
that the absolute value of \MLK\ is not only IMF--dependent but also
depends on the specific stellar population synthesis model used
(because of the differing treatment of stellar remnants) and should
therefore be taken with caution.  Instead, we concentrate on the
relative change of \MLK\, and therefore the relative change of the MF,
with $z$.

Star formation in our $K$-selected objects begins well before $z \sim
1.5$. The bulk of the stars in the local universe is therefore older
than 8~Gyr and formed before $z = 1$, consistent with recent estimates
of the total stellar mass density of the universe, which suggest that
the stellar mass density at $z \sim 1$ is roughly half its local value
(see also Sect.~\ref{sec:mass-density};
\citealp{DPFB03,Fontanaetal03,2dF01}).  Only very few objects at low
redshift have $\MLK < 0.5$ and therefore mean ages less than $\sim
3$~Gyr (see Fig.~\ref{fig:mlk-csps}). This is no surprise in a K--band
selected survey with a magnitude limit of $K < 19.5$. There are very
few objects (well below 1\% of the sample) that scatter to ages above
the age of the universe despite the age prior included in the
likelihood function. Inspection of those objects shows that they are
most likely photometrically problematic objects (blended objects;
objects with strong emission lines; objects affected by bright star
artifacts; extended low--$z$ objects with an overlapping foreground
star).  The large majority of objetcts, however, yields acceptable
($\chi^2 < 10$) fits to the model grid.

As expected, the range of \ML\ values evolves steadily toward lower
averages with increasing redshift as the stars inevitably become
younger. No population of very young objects with ages of $\sim 1$~Gyr
is found at $z < 1$, although there might be such populations at
magnitudes below our detection limit of $K < 19.5$ and hence stellar
masses below $\sim 10^{10}$~\hMsun\ at lower redshift.  This again
indicates that the objects in our sample experienced their most
intensive phase of star formation at redshifts above $z \sim 1.5$.
Indeed, closer to the redshift limit of the current analysis at $z =
1.2$, objects with ages of around 1~Gyr do appear, while older objects
with ages of 3~Gyr are still present up to the survey's redshift
limit.  This indicates that we approach the epoch of the formation of
some of these systems although we do not quite reach it with MUNICS.
The presence of apparently old and massive stellar systems at $z \sim
1$ has also been pointed out recently by several authors studying
spectroscopic samples of extremely red objects (EROs), e.g., by the
K20 survey \citep{K20-02} using deep optical spectroscopy, and by
\citet{Saraccoetal03}, by spectroscopic followup of MUNICS--selected
EROs in the NIR. It will be very interesting to see the results from
larger and deeper surveys targeted at galaxies in the redshift range
$1 < z < 2$.

As is observed in the local universe, lower mass galaxies have on
average lower \ML\ values at all redshifts in our sample.  This is
shown in Fig.~\ref{f:mean-mlk-mass} and Table~\ref{tab:mean-mlk-mass}.
We wish to point out, however, that the evolution of \MLK\ with
redshift seems quite independent of galaxy stellar mass. If anything,
more massive systems show a slightly flatter evolution. In addition,
older systems have lower dust extinction, and there are no strongly
dust reddened objects with $A_V > 1$ in the sample.

Finally, we construct the stellar MF, using the \Vmax\ method to
account for the fact that fainter galaxies are not visible in the
whole survey volume. Here each galaxy in a given redshift bin
$[z_l,z_h)$ contributes to the number density an amount inversely
proportional to the volume in which the galaxy is detectable in this
redshift bin. The method is fully analogous to the one used for the LF
in MUNICS~II, and the reader is referred to that work for details.

\begin{figure*}
  \centering 
  \epsscale{0.8}
  \plotone{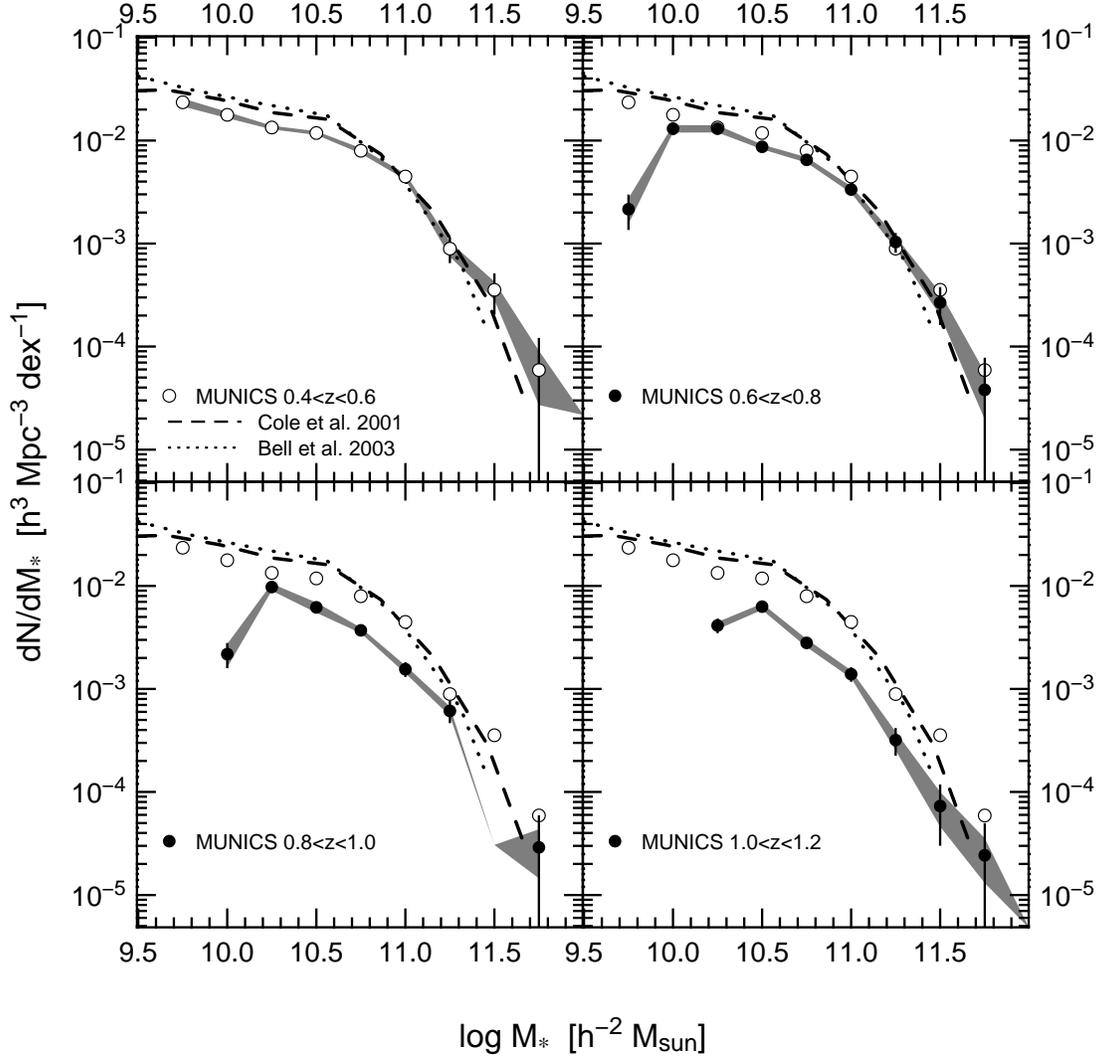}
  \caption{\label{f:mf}%
    The evolution of the stellar MF with redshift. The open
    symbols are the MUNICS values at $0.4 < z < 0.6$, the closed
    symbols are the MUNICS values at higher redshifts. The lowest $z$
    values are shown in all panels for comparison. Error bars denote
    the uncertainty due to Poisson statistics. The shaded areas show
    the 1~$\sigma$ range of variation in the MF given the
    total systematic uncertainty in \MLK\ discussed in
    Sect.~\ref{sec:deriving-masses}. The dotted and dashed lines show
    the $z=0$ stellar MF derived similarly to our methods
    using SDSS, 2dF, and 2MASS data  \citep{BMKW03,2dF01}.}
\end{figure*}

Fig.~\ref{f:mf} shows the MF of galaxies with $10^{10} < M/(\hMsun) <
10^{12}$ in four redshift bins centered at $z=0.5$, $z=0.7$, $z=0.9$,
and $z=1.1$. The results for the local stellar MF from \citet{2dF01}
and \citet{BMKW03}, derived by fitting stellar population models to
multicolor photometry and deriving NIR \ML\ values
similarly to the method employed here are also shown for comparison as
a dashed and a dotted line, respectively. The curve from \citet{BMKW03}
has been corrected to a 30\% higher normalization to account for their
choice of IMF, which is a Salpeter form with a lower fraction of low
mass stars (``diet''-IMF) that yields 30\% lower masses.  The data from
the lowest redshift bin, $0.4 < z < 0.6$, are shown alongside the
higher--redshift data, for easier comparison, as open symbols. Error
bars denote the uncertainty due to Poisson statistics. The shaded
areas show the 1~$\sigma$ range of variation in the MF from
Monte--Carlo simulations given the total systematic uncertainty in
\MLK\ of 30\% discussed in Sect.~\ref{sec:deriving-masses}, assuming a
Gaussian distribution.  Although this is not strictly realistic, we
note that systematic uncertainties discussed above are not dominated
by any one particular asymmetric bias. The lost--light effect is not
included in these simulations. The values of our MFs are given in
Table~\ref{tab:mf-results}, along with the errors from Poisson
statistics and the systematic uncertainties in \MLK. This table also
quotes the values we obtain assuming that all galaxies follow an
$r^{1/4}$ light profile, this case being the one with the largest
corrections to the Kron--like photometry.

We stress the fact that the highest mass bin, at $\log M/(\hMsun) =
11.75$, typically contains only one object and the next bin, at $\log
M/(\hMsun) = 11.5$, typically contains fewer than five objects. Hence,
moving objects from one bin to the other at the very bright end
(either through the uncertainty in \MLK\ or through lost--light
corrections to the photometry) has a big effect. Furthermore, the MF
is steepest here, so a small change to the luminosity has a big effect
on the derived mass. Therefore, the uncertainties at the very bright
end are much larger than those around the knee of the MF (around the
characteristic mass, $M^*$, where bins contain typically $\sim 200$
objects). The mass bin $\log M/(\hMsun) = 11.5$ at z=0.9 contains no
objects, and therefore the uncertainty contours are artificially
compressed there. At the very bright end, the Poisson errors dominate
our total uncertainty in the MF. Shot noise and \MLK\ systematics
contribute equally at the knee around $M^*$. The lost--light
uncertainties are smaller than the shot noise and the systematics
associated with assigning \MLK.

The lowest redshift bin shows remarkable agreement with the $z=0$
values, despite the different selection at low versus high redshift
and the different model grids used, although we obtain slightly lower
number densities at $\log M/(\hMsun) \leq 10.5$. Therefore, there
seems to be not much evolution in stellar mass at $z < 0.5$. The
general trend at higher redshift is for the total normalization of the
stellar MF to go down and for the knee to move toward
lower masses. This causes the higher masses to evolve faster in number
density than lower masses, is well visible in Fig.~\ref{f:mf} at
$10.5 < \log M/(\hMsun) < 11.5$ and is further discussed in
Sect.~\ref{sec:num-density}.

These results have to be seen in the context of the evolution of the
K--band LF (MUNICS~II, using the same sample as in this work; also
\citealp{K20-03,LCIRS02}). It is important to stress that the results
obtained here depend strongly on the quality of the underlying LF. The
trends with mass observed here depend on the exact shape of the LF
around the characteristic luminosity, $L^*$, and its evolution with
$z$. Since the relative change in \MLK\ is similar at all masses
(Fig.~\ref{f:mean-mlk-mass}), the increase in number density evolution
with mass can be explained in the following way: as one is moving to
higher masses (at any given \MLK), the corresponding luminosity is
moving down the steepening part of the LF, so that the same relative
change in \MLK\ yields a higher change in the number density at a
higher stellar mass.  This is a very fundamental observation and it is
hard to see how this can be avoided.


\section{The number density of massive systems}\label{sec:num-density}

In this section we concentrate on the number density of galaxies
having stellar masses exceeding some mass limits, in other words, the
integrated stellar MF. Although this information is already implicitly
contained in Fig.~\ref{f:mf}, its integrated form is less noisy and
has been
frequently discussed in recent literature (e.g.\ MUNICS~III; \citealp{%
Imetal96,KCW96,Driveretal98,TY98,KC98a,Fontanaetal99,LCIRS02,Cohen02}).

\begin{figure*}
  \centering 
  \epsscale{1.0}
  \plotone{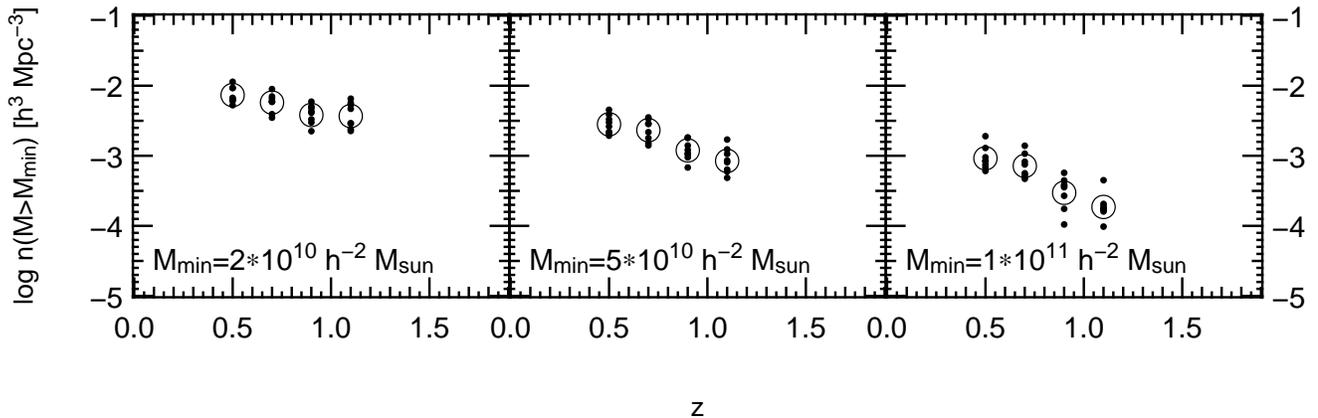}
  \caption{\label{f:intmf}%
    Co-moving number density of objects having stellar masses
    exceeding $\Mlim = 2\times 10^{10}\,\hMsun$, $\Mlim = 5\times
    10^{10}\,\hMsun$, and $\Mlim = 1\times 10^{11}\,\hMsun$ The solid
    points denote the values measured separately in each survey field,
    the open circles denote the mean values over the whole survey
    area.  values.}
\end{figure*}

We present the results in the same fashion that was used in
MUNICS~III, where we used the assumption that all stars form at
$z=\infty$ to maximize the stellar mass at any K--band luminosity.  We
plot the number density of systems more massive than a given mass
\Mlim. We use $\Mlim = 2\times 10^{10}\,\hMsun$, $\Mlim = 5\times
10^{10}\, \hMsun$, and $\Mlim = 1\times 10^{11}\,\hMsun$. The results
are shown in Fig.~\ref{f:intmf} and listed in Table~\ref{tab:intmf}.

Again, it is clearly visible that objects with higher stellar masses
show stronger evolution in number density, a result that we have
already seen in the stellar MF in Sect.~\ref{sec:mass-function}. The
number density of objects more massive than $2\times 10^{10}\,\hMsun$
declines by a factor of $\sim 2$ to $z \sim 1$, again pointing to half
the stellar mass having formed since $z \sim 1$ (see discussion
below). For objects having $M > 5\times 10^{10}\,\hMsun$, the decline
is by a factor of 3.3. Objects with $M>10^{11}\,\hMsun$ evolve by a
factor of 4.9 in number density.  If we stretch all sources of
uncertainty to their maximum, including the lost--light corrections,
this number changes to a factor of 2.7.  For stellar masses below
$10^{11}\,\hMsun$, the uncertainties play a much smaller role.

We use Fig.~\ref{f:intmf} to make another point concerning cosmic
variance. The figure shows the average number densities in our survey,
but also the individual numbers from eight disjoint survey patches,
each approximately 130 arcmin$^2$\ in size (see MUNICS~II). There is
considerable variance among the survey fields even for the lowest mass
limit where the number of objects is largest. It is also apparent,
that the variance increases rapidly toward higher limiting mass, not
only because of smaller numbers, but also because of the higher
clustering of massive galaxies (e.g.\ 
\citealp{DCetal00,2dF01a,2dF03}).


\section{The total stellar mass density}\label{sec:mass-density}

\begin{figure}
  \centering 
  \epsscale{0.7}
  \plotone{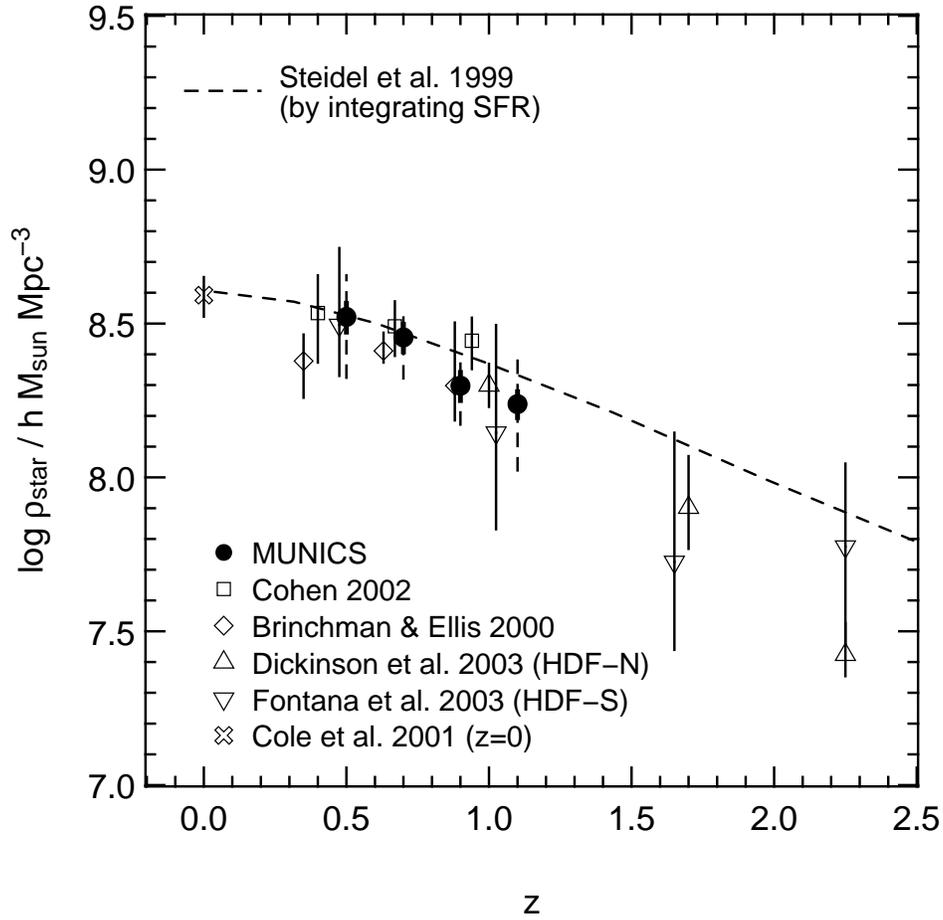}
  \caption{\label{f:md}%
    The evolution of the total stellar mass density in the universe.
    The closed circles are the MUNICS values, open symbols are values
    from the literature. The integrated star formation rate (dashed
    curve) is shown for comparison. The thick error bars on the MUNICS
    values (solid dots) are the statistical errors associated with the
    data. The dashed error bars show the variance we get from the
    MUNICS data in GOODS size patches.}
\end{figure}

The total stellar mass density of the universe, observationally the
complement of the SFH of the universe (e.g.,
\citealp{Madauetal96,SAGDP99}), is shown in Fig.~\ref{f:md}. We also
plot the local value from \citet{2dF01}, values from the HDFs
\citep{DPFB03,Fontanaetal03} covering $z > 1$, and values from
\citet{Cohen02} and \citet{BE00} at $z < 1$. Additionally, we
integrate the SFH curve (including extinction correction) from
\citet{SAGDP99} for comparison. We also list the results in
Table~\ref{tab:mass-dens}.

At $z >= 0.9$ we do not sample the stellar MF to low enough
masses to be able to compute the necessary integral without
corrections. We therefore assume that the faint end slope does not
change with redshift, using the faint end slope of a Schechter fit to
our $z=0.5$ data normalized to the higher redshift MFs to
compute corrections to the total stellar mass. This correction is a
factor of 1.35 in the highest redshift bin.

The results from these different surveys agree reasonably well, given
their differing selection techniques and methodologies. While
\citet{BE00} used a small (321 objects) sample and a method similar to
ours to derive \MLK\, \citet{Cohen02} uses galaxy SED evolutionary
models (assigning fixed SFHs and formation redshifts to $z=0$ galaxy
types) to read off the change in \MLK\ with $z$. This is then used to
convert K--band luminosity densities to stellar mass densities,
assuming $\MLK\ = 0.8$ for all galaxy types locally. This method
typically yields higher \ML\ values than actual SED fitting where the
star formation rate and mean age at each redshift are essentially free
parameters and are allowed to produce younger objects as needed. We
can compare our average B--band \ML\ to the numbers obtained by
\citet{DPFB03}. They quote average \MLB\ values in the redshift range
0.5 -- 1.4 between 0.96 and 1.38 depending on the model (number of
components and metallicity) which compares well to our average values
from Table~\ref{tab:mean-mlk-mass}. The average \MLB\ for our whole
sample is 1.28.

From our data it appears that 50\% of the local mass in stars has
formed since $z \sim 1$. This is consistent with results obtained in
the HDFs (\citealp{DPFB03,Fontanaetal03,Rudnicketal03}), although the
volume probed by the HDFs at $z < 1$ is very small and the results
therefore uncertain at these redshifts. The results for the integrated
star formation rate of the universe traced by the UV continuum
emission agree with our results as well as the HDF results, reasonably
well.  The data from \citet{BE00} are consistent with ours at $z \sim
1$ but are lower at $z \sim 0.4$ and seem to under--predict the local
value if extrapolated.  The values obtained by \citet{Cohen02} are
higher at $z \sim 1$, which we think is attributable to their method of
deriving \MLK.

It is worth noting that the results from the HDF--N differ by a factor
$\sim 2$ from the HDF--S results, which is attributable to cosmic
variance.  The MUNICS values in Fig.~\ref{f:md} are shown with their
statistical errors (thick error bars), which amount to roughly 10\%.
We also show the variance we get in our sample divided into GOODS size
patches of 150 arcmin$^2$\ (dashed error bars), showing that at these
redshifts, even surveys like GOODS (Great Observatories Origins Deep
Survey) are expected to be dominated by cosmic variance. We expect
differences of around 50\% between the two GOODS areas.


\section{Discussion and conclusions}\label{sec:discussion}

The results of this study of the evolution of stellar masses of
galaxies to $z = 1.2$ can be summarized as follows.

1. The mass--to--light ratios (\ML s) in the K--band decline with
  redshift by similar amounts for all stellar masses above
  $10^{10}\,\hMsun$.  Lower mass galaxies have lower \ML\ values at
  all redshifts.

2. The stellar mass function evolves significantly to $z = 1.2$.
  The total normalization decreases by a factor of $\sim 2$, the
  characteristic mass (the knee) shifts towards lower masses and the
  bright end therefore steepens with redshift.

3. The amount of number density evolution is a strong function of
  stellar mass. More massive systems show stronger evolution in their
  number density.

4. In total, roughly half the stellar mass in the present day
  universe forms at $0 < z < 1$.

At first sight these results seem to disagree with recent measurements
of the number density of morphologically selected or color--selected
(extremely red objects, or EROs) early--type systems
(\citealp{Imetal96,HSTCFRS399,MCD00,DCetal00a}; but see also
\citealp{TS99,MEABC99}). However, our sample is K--band flux limited
and therefore largely morphologically blind. The statement that the
number densities of galaxies as a function of their stellar mass
content evolves does not contradict a constant number density of
elliptical galaxies, as long as their morphologies do not change (on
average) as their stellar mass grows. Early type galaxies in clusters,
however, show clear evidence for passive evolution, at least up to $z
\sim 1$ (e.g., \citealp{BZB96,ZB97,SED98,KIDF00c,KIFD01,DF01,DS03})
which implies a constant stellar mass. In the field
\citep{DFKI01,Treuetal01c}, early--type galaxies seem to show larger
age scatter, and in a recent study of a sample of strong gravitational
lens galaxies, \citet{VDF03} found that about half of the field
elliptical galaxies have younger stellar populations and are best
fitted by formation redshifts extending down to $z_f \sim 1$, a
scenario that is fully compatible with our results.

Furthermore, we still find that more massive galaxies have {\em older}
luminosity weighted mean ages (higher \ML) and that while galaxies
with young populations are present in larger numbers at $z \sim 1$,
almost maximally old systems are still present even at the highest
redshift we probe. This has also been demonstrated by spectroscopic
studies of samples of EROs, most recently by the K20 survey
\citep{K20-02}, using deep optical spectroscopy, and by
\citet{Saraccoetal03}, using spectroscopic follow-up of
MUNICS--selected EROs in the NIR.

While half the present day stellar mass seems to form after $z \sim
1$, it seems that the oldest stellar populations at each epoch are
harbored by the most massive objects. The growth of stellar mass at
the high mass end of the stellar MF is driven by accretion and merging
of objects having similarly old populations, rather than by star
formation.  This must be the case because if star formation were to
contribute equally to the growth of stellar mass at all masses, the
ratio of young to old stars, and hence \ML, would be the same in all
galaxies, which is not the case. The higher \ML\ of the more massive
galaxies indicates that star formation since $z \sim 1$ is not as
important in these objects as it is at lower masses. A similar result
in the local universe was shown by \citet{KTC94}, and \citet{BE00}
explicitly arrived at the same conclusion by looking at the star
formation rate per unit stellar mass in their high-$z$ sample.

To account for their rapid change in number density, however, merging
and accretion have to dominate. Indeed, \citet{CBDP03} find that in
the HDF, the merger fraction of massive galaxies increases more
rapidly than that of less massive galaxies. Another argument in
support of this picture has been put forward by \citet{CSHC96}. By
combining K--band imaging and star formation measurements from OII
equivalent widths, they showed that the star formation rate per unit K
band luminosity (as a surrogate for stellar mass) is lower in higher
mass systems and that this quantity increases with redshift, the
increase being slower for higher mass systems (i.e., they formed
earlier). In addition, many authors noted a steep increase in the star
formation rate at $0 < z < 1$, (e.g., 
\citealp{CFRS96,CFRS1297,RRetal97,Floresetal99}).

We wish to point out that the observed trend in density evolution as a
function of mass is qualitatively consistent with the expectation from
hierarchical galaxy formation models. The most rapid evolution is
predicted for the number density of the most massive galaxies, while
the number density of less massive galaxies is predicted to evolve
less. While older models tended to over--predict this evolution, more
recent models seem to yield results more consistent with the redshift
distribution of K--band selected galaxies
\citep{Fontanaetal99,LCIRS02}. A detailed comparison of the predicted
stellar MFs and colors of galaxies with these observations has yet to
be done.

Upcoming large area surveys will advance this field significantly by
providing more colors, morphological information, and spectroscopic
information. This will allow us to use better CSP model grids, divide
the samples up by morphology, and include direct star formation
measurements from spectroscopy. This will allow us to determine the
relative contributions of star formation and accretion/merging to the
mass build--up as a function of cosmic epoch and galaxy properties.
Hence these datasets will provide a much more complete picture of the
assembly process of galaxies.


\acknowledgments

This work was partly supported by the Deutsche Forschungsgemeinschaft,
grant SFB 375 ``Astroteilchenphysik.'' N.D.\ acknowledges support by
the Alexander von Humboldt Foundation.  We would also like to thank
the Calar Alto staff for their long--standing support during many
observing runs over the last six years. We would like to thank Jarle
Brinchmann for his careful reading of the manuscript.

This work is based on observations obtained at (1) the German--Spanish
Astronomical Center, Calar Alto, operated by the Max--Planck Institut
f\"ur Astronomie, Heidelberg, jointly with the Spanish National
Commission for Astronomy, (2) McDonald Observatory, operated by the
University of Texas at Austin, and the Hobby--Eberly Telescope,
operated by McDonald Observatory on behalf of The University of Texas
at Austin, the Pennsylvania State University, Stanford University,
Ludwig--Maximilians--Universit\"at M\"unchen, and
Georg--August--Universit\"at G\"ottingen, and (3) the European
Southern Observatory, Chile, proposals 66.A-0129 and 66.A-0123.



\onecolumn

\clearpage

\begin{deluxetable}{ccccccc}
  \centering
  \tablewidth{0pt}
  \tabletypesize{\small}  
  \tablecaption{\label{tab:mean-mlk-mass}%
    Average K--band and B--band \ML\ value as a function of
    redshift.}
  \tablecolumns{7}
  \tablehead{%
    \colhead{$z$} & 
    \multicolumn{2}{c}{$2\E{10} < M/(\hMsun) < 5\E{10}$} &
    \multicolumn{2}{c}{$5\E{10} < M/(\hMsun) < 1\E{11}$} &
    \multicolumn{2}{c}{$M/(\hMsun) > 1\E{11}$} \\ 
    & \colhead{$\langle \MLK\ \rangle$} & 
    \colhead{$\langle \MLB\ \rangle$} & 
    \colhead{$\langle \MLK\ \rangle$} & 
    \colhead{$\langle \MLB\ \rangle$} & 
    \colhead{$\langle \MLK\ \rangle$} & 
    \colhead{$\langle \MLB\ \rangle$}}
  \startdata
          0.5 & 0.65 & 1.69 & 0.71 & 2.18 & 0.74 & 2.58\\ 
          0.7 & 0.58 & 1.36 & 0.64 & 1.88 & 0.68 & 2.27\\ 
          0.9 & 0.46 & 0.61 & 0.55 & 1.26 & 0.62 & 1.83\\ 
          1.1 & 0.35 & 0.23 & 0.43 & 0.49 & 0.55 & 1.23\\
 \enddata
\end{deluxetable}

\begin{deluxetable}{lllll|lllll}
  \centering
  \tablewidth{0pt}
  \tabletypesize{\scriptsize}
  \tablecaption{\label{tab:mf-results}%
    The values of the stellar mass function in 4 redshift bins.}
  \tablecolumns{10}
  \tablehead{%
    \colhead{$z$} &
    \colhead{$\log M$} & 
    \colhead{$dN/dM$} & 
    \colhead{\p $\sqrt(N)$ \p \MLK\ \tablenotemark{a}} & 
    \colhead{max ($r^{1/4}$) \tablenotemark{b}} & 
    \colhead{$z$} &
    \colhead{$\log M$} & 
    \colhead{$dN/dM$} & 
    \colhead{\p $\sqrt(N)$ \p \MLK\ \tablenotemark{a}} & 
    \colhead{max ($r^{1/4}$) \tablenotemark{b}} \\ 
    & \colhead{$\hMsun$} & 
    \colhead{$h^3~\Mpc^{-3}$} &
    \colhead{$h^3~\Mpc^{-3}$} & 
    \colhead{$h^3~\Mpc^{-3}$} &
    & \colhead{$\hMsun$} & 
    \colhead{$h^3~\Mpc^{-3}$} & 
    \colhead{$h^3~\Mpc^{-3}$} & 
    \colhead{$h^3~\Mpc^{-3}$}}
    \startdata
    0.5 & 9.75 & 2.35\E{-2} & (\p 2.29\p 2.07)\E{-3} & 2.34\E{-2} & 
    0.7 & 9.75 & \nodata & \nodata & \nodata \\ 
    & 10.00 & 1.78\E{-2} & (\p 1.27\p 1.24)\E{-3} & 1.74\E{-2} & 
    & 10.00 & 1.32\E{-2} & (\p 1.46\p 1.08)\E{-3} & 1.02\E{-2}\\ 
    & 10.25 & 1.34\E{-2} & (\p 9.06\p 5.36)\E{-4} & 1.26\E{-2} & 
    & 10.25 & 1.30\E{-2} & (\p 0.88\p 1.06)\E{-3} & 1.34\E{-2}\\ 
    & 10.50 & 1.18\E{-2} & (\p 8.43\p 4.80)\E{-4} & 1.06\E{-2} & 
    & 10.50 & 8.68\E{-3} & (\p 5.91\p 4.56)\E{-4} & 8.17\E{-3}\\ 
    & 10.75 & 7.95\E{-3} & (\p 6.90\p 4.77)\E{-4} & 7.31\E{-3} & 
    & 10.75 & 6.49\E{-3} & (\p 5.01\p 3.94)\E{-4} & 6.16\E{-3}\\
    & 11.00 & 4.47\E{-3} & (\p 5.17\p 2.94)\E{-4} & 5.61\E{-3} & 
    & 11.00 & 3.35\E{-3} & (\p 3.59\p 2.51)\E{-4} & 3.70\E{-3}\\ 
    & 11.25 & 8.92\E{-4} & (\p 2.30\p 1.48)\E{-4} & 2.56\E{-3} & 
    & 11.25 & 1.03\E{-3} & (\p 1.99\p 1.38)\E{-4} & 1.92\E{-3}\\ 
    & 11.50 & 3.55\E{-4} & (\p 1.45\p 0.72)\E{-4} & 4.15\E{-4} & 
    & 11.50 & 2.67\E{-4} & (\p 1.01\p 0.67)\E{-4} & 7.27\E{-4}\\ 
    & 11.75 & 5.92\E{-5} & (\p 5.92\p 3.21)\E{-5} & 5.92\E{-5} & 
    & 11.75 & 3.80\E{-5} & (\p 3.80\p 1.84)\E{-5} & 2.29\E{-4}\\ 
    \tableline 
    0.9 & 10.25 & 9.76\E{-3} & (\p 9.39\p 7.66)\E{-4} & 7.39\E{-3} & 
    1.1 & 10.25 & \nodata & \nodata & \nodata \\ 
    & 10.50 & 6.19\E{-3} & (\p 5.24\p 7.12)\E{-4} & 6.17\E{-3} & 
    & 10.50 & 6.23\E{-3} & (\p 5.61\p 4.51)\E{-4} & 5.40\E{-3}\\ 
    & 10.75 & 3.71\E{-3} & (\p 3.32\p 2.26)\E{-4} & 3.31\E{-3} & 
    & 10.75 & 2.80\E{-3} & (\p 2.80\p 2.53)\E{-4} & 3.24\E{-3}\\ 
    & 11.00 & 1.55\E{-3} & (\p 2.14\p 1.33)\E{-4} & 2.41\E{-3} & 
    & 11.00 & 1.39\E{-3} & (\p 1.88\p 1.20)\E{-4} & 1.72\E{-3}\\ 
    & 11.25 & 6.12\E{-4} & (\p 1.34\p 0.64)\E{-4} & 7.59\E{-4} & 
    & 11.25 & 3.19\E{-4} & (\p 8.86\p 6.55)\E{-5} & 6.87\E{-4}\\ 
    & 11.50 & \nodata & \nodata & 3.21\E{-4} & 
    & 11.50 & 7.31\E{-5} & (\p 4.22\p 2.70)\E{-5} & 1.46\E{-4}\\ 
    & 11.75 & 2.90\E{-5} & (\p 2.90\p 1.45)\E{-5} & 5.79\E{-5} & 
    & 11.75 & 2.43\E{-5} & (\p 2.43\p 1.13)\E{-5} & 4.91\E{-5}\\ 
    \enddata
  \tablenotetext{a}{Poisson errors and systematic uncertainty in \MLK.}  
  \tablenotetext{b}{Maximal values usinglost--light corrections assuming 
    de Vaucouleurs profiles.}
\end{deluxetable}

\begin{deluxetable}{cccc}
  \centering
  \tablewidth{0pt}
  \tablecaption{\label{tab:intmf}%
    Number density of objects with stellar mass $M > \Mlim$.}
  \tablecolumns{4}
  \tablehead{%
    \colhead{$z$} & 
    \colhead{$n(M>2\E{10}\,\hMsun)$} & 
    \colhead{$n(M>5\E{10}\,\hMsun)$} &
    \colhead{$n(M>1\E{11}\,\hMsun)$}\\ 
    & \colhead{$h^3\,\Mpc^{-3}$} &
    \colhead{$h^3\,\Mpc^{-3}$} &
    \colhead{$h^3\,\Mpc^{-3}$}}
  \startdata
  0.5 &(7.37\p 0.95)\E{-03}&(2.82\p 0.59)\E{-03}&(9.22\p 3.38)\E{-04}\\
  0.7 &(5.76\p 0.68)\E{-03}&(2.32\p 0.43)\E{-03}&(7.18\p 2.37)\E{-04}\\ 
  0.9 &(3.83\p 0.58)\E{-03}&(1.19\p 0.27)\E{-03}&(2.97\p 1.37)\E{-04}\\ 
  1.1 &(3.70\p 0.49)\E{-03}&(8.47\p 2.16)\E{-04}&(1.86\p 0.98)\E{-04}\\
  \enddata
\end{deluxetable}

\begin{deluxetable}{cccc}
  \centering
  \tablewidth{0pt}
  \tablecaption{\label{tab:mass-dens}%
    The total stellar mass density of the universe.}
  \tablecolumns{4}
  \tablehead{%
    \colhead{$z$} & 
    \colhead{$\log \rhos$} & 
    \colhead{$\delta \log \rhos$} & 
    \colhead{$\delta^{\mathrm{GOODS}} \log \rhos$}\\ 
    & \multicolumn{3}{c}{$h^{-3}\, \Msun\, \Mpc^{-3}$}}
  \startdata
  0.5 & 8.52 & 0.044 & 0.14\\ 
  0.7 & 8.45 & 0.042 & 0.13\\ 
  0.9 & 8.29 & 0.043 & 0.12\\ 
  1.1 & 8.24 & 0.040 & 0.14\\
  \enddata
\end{deluxetable}

\end{document}